\documentclass[preprint]{elsarticle}

\usepackage{hyperref}
\journal{Journal of Magnetism and Magnetic Materials}

\begin{document}

\begin{frontmatter}

\title{Phase diagrams of a 2D Ising spin-pseudospin model}

\author[mymainaddress,mycorrespondingauthor]{Yu.D. Panov}
\cortext[mycorrespondingauthor]{Corresponding author}
\ead{yuri.panov@urfu.ru}

\author[mymainaddress]{V.A. Ulitko}
\author[mymainaddress]{K.S. Budrin}
\author[mymainaddress]{A.A. Chikov}
\author[mymainaddress]{A.S. Moskvin}
\address[mymainaddress]{Ural Federal University, Ekaterinburg, 620002, Russia}

\begin{abstract}
We consider the competition of magnetic and charge ordering in high-Tc cuprates within the framework of the simplified static 2D spin-pseudospin model. 
This model is equivalent to the 2D dilute antiferromagnetic (AFM) Ising model with charged impurities. 
We present the mean-field results for the system under study 
and make a brief comparison with classical Monte Carlo (MC) calculations.
Numerical simulations show that the cases of strong exchange and strong charge correlation differ qualitatively. 
For a strong exchange, the AFM phase is unstable with respect to the phase separation (PS) into the charge and spin subsystems, which behave like immiscible quantum liquids. 
An analytical expression was obtained for the PS temperature.
\end{abstract}


\end{frontmatter}

\section{Introduction}
One of the topical problems in the physics of high-T$_c$ cuprates is the coexistence and competition of the spin, superconducting, and charge orderings. 
The studying of the interplay between magnetism and superconductivity 
in cuprates has a long history~\cite{Birgeneau2006,Tranquada2014jmmm,Tranquada2014physb}.  
Over the last fifteen years, a wealth of experimental results has suggested the presence of the charge 
ordering~\cite{Wu2011, Ghiringhelli2012,Chang2012,LeBoeuf2013,Torchinsky2013,Park2014,DaSilvaNeto2014,Sonier2015,Wu2015,Comin2016,Laliberte2018}
and the interplaying spin and charge orderings~\cite{Abbamonte2005,Berg2009,Fujita2012,Fradkin2012,Croft2014,Drachuck2014,Cyr-Choiniere2015} 
in cuprates.
Recently \cite{MoskvinPRB2011} we argued that an unique property of high-T$_c$ cuprates is related to the charge-transfer instability in the CuO$_2$ planes. 
For the CuO$_4$ centers in CuO$_2$ plane, 
this implies accounting of the three many-electron valence states CuO$_4^{5-,6-,7-}$ (nominally Cu$^{1+;2+;3+}$) 
as the components of the $S = 1$ pseudospin triplet with $M_S = -1,0,+1$, respectively,
and allows us to use of the $S=1$ pseudospin formalism 
\cite{MoskvinPRB2011,MoskvinJPhys2013}.
To consider the competition of spin and charge orderings in cuprates, a simplified static 2D spin-pseudospin model was proposed \cite{PanovJSNM2016,PanovJLTP2017,PanovJETPl2017} as a limiting case of a general pseudospin model.

The Hamiltonian of the static spin-pseudospin model is:
\begin{equation}
		{\mathcal H} = 
		\Delta \sum_i S_{zi}^2 
		+ V \sum_{\left\langle ij\right\rangle} S_{zi} S_{zj} 
		+ \tilde{J} \sum_{\left\langle ij\right\rangle} \sigma_{zi} \sigma_{zj}
		- \tilde{h} \sum_i \sigma_{zi} 
		- \mu \sum_i S_{zi}
		,
	\label{H}
\end{equation}
where $S_{zi}$ is a $z$-projection of the on-site pseudospin $S=1$ 
and $\sigma_{zi}=P_{0i}s_{zi}/s$  is a normalized $z$-projection of conventional spin $s=1/2$ operator, 
multiplied  by the projection operator $P_{0i} = 1-S_{iz}^2$.
The $\Delta=U/2$ is the on-site correlation and $V>0$ is the inter-site density-density interaction, 
$J=\tilde{J}/s^2>0$ is the Cu$^{2+}{-}$Cu$^{2+}$ Ising spin exchange coupling, 
$h=\tilde{h}/s$ is the external magnetic field, 
$\mu$ is the chemical potential, 
so we assume the total charge constraint, 
$nN = \sum \left\langle S_{zi} \right\rangle = const$, 
where $n$ is the density of doped charge.
The sums run over the sites of a 2D square lattice,
$\left\langle ij \right\rangle$ means the nearest neighbors.
This spin-pseudospin model generalizes the 2D dilute AFM Ising model with charged impurities. 
In the limit $\Delta\rightarrow-\infty$ it reduces to the $S=\frac{1}{2}$ Ising model with fixed magnetization. 
At $\Delta>0$ the results can be compared with the Blume--Capel model~\cite{Blume1966,Capel1966,Capel1967} 
or with the  Blume--Emery--Griffiths model~\cite{Blume1971}.

An analysis of the ground state (GS) phase diagrams was done within the mean field approach \cite{PanovJSNM2016,PanovJLTP2017}.  
It was shown, that the five GS phases are realized in two limits. 
In a weak exchange limit, at $\tilde{J} < V$, all the GS phases (COI, COII, COIII, FIM) correspond to the charge ordering (CO) of a checker-board type at mean charge density $n$. 
While the COI phase is the charge-ordered one without spin centers,  
the COII and COIII phases are diluted by the non-interacting spins distributed in one sublattice only. 
This ferrimagnetic spin ordering is a result of the mean-field approach, so the classical MC calculations show a paramagnetic response at low temperatures.
The FIM phase is also formally ferrimagnetic.
Here the AFM spin ordering is diluted by the non-interacting charges distributed in one sublattice. 
In a strong exchange limit, at $\tilde{J} > V$, there are only COI phase and AFM phase with the charges distributed in both sublattices.
The absence of charge transfer in the Hamiltonian~(\ref{H}) is the most important limitation of our present model 
for comparison with the actual phase diagram of cuprates. 
But, as shown in~\cite{Panov2018}, the accounting for two-particle transport enriches the GS phase diagram of the spin-pseudospin model with superfluid and supersolid phases competing with CO phases.

The paper is organized as follows. 
We present the mean-field results for the system under study 
and make a brief comparison with the MC calculations in section 2.  
The MC calculations show, that in a strong exchange case 
the AFM phase is unstable with respect to the PS into the charge and spin subsystems. 
In section 3 we analyse the thermodynamic properties of the phase separation (PS) state in a framework of coexistence of two homogeneous phases.  
Finally, section 4 is devoted to conclusions.

\section{Mean-field approximation}

Here we outline briefly the results for the system under study in the mean-field approximation (MFA). 
We use the Bogolyubov inequality~\cite{Kuzemsky2015} for the grand potential $\Omega({\mathcal H})$: 
$\Omega({\mathcal H}) \;\leq\; \Omega = \Omega({\mathcal H}_0) + \langle {\mathcal H} - {\mathcal H}_0 \rangle$. 
In the standard way, we divide the square lattice into two sublattices, $A$ and $B$, and choose
\begin{equation}
	\beta {\mathcal H}_0 =
		\delta \sum_i S_{zi}^2
	- \sum_{\alpha,i_\alpha} \beta_\alpha S_{z i_\alpha}
	- \sum_{\alpha,i_\alpha} \gamma_\alpha \sigma_{z i_\alpha}
	,
\end{equation}
where $\beta=1/T$, $\delta = \beta\Delta$, $\beta_\alpha$ and $\gamma_\alpha$ are the molecular fields, $\alpha=A,B$. 
For the estimate $\omega=\Omega/N$ we get
\begin{eqnarray}
	2\beta \, \omega 
	&=& 
	\sum_{\alpha} 
	\Big[ 
	\left( \beta_\alpha - \xi \right)  S_\alpha
	+ \left( \gamma_\alpha - \eta \right) \sigma_\alpha
	- \ln 2 \left( e^{-\delta} \cosh \beta_\alpha + \cosh \gamma_\alpha \right) 
	\Big] + {}
	\nonumber \\
	&&
	{}+ z \nu \, S_A S_B + z j \, \sigma_A \sigma_B
	, 
\end{eqnarray}
where $\xi = \beta \mu$, $\nu = \beta V$, $j = \beta\tilde{J}$, $\eta = \beta \tilde{h}$, 
$z=4$ is the number of nearest neighbours,
and the average sublattice (pseudo)magnetizations
$\left\langle S_z \right\rangle_\alpha = S_\alpha$ and $\left\langle \sigma_z \right\rangle_\alpha = \sigma_\alpha$ have the form
\begin{equation}
	S_\alpha = \frac{ \sinh \beta_\alpha}{\cosh \beta_\alpha + e^{\delta} \cosh \gamma_\alpha}
	,\qquad
	\sigma_\alpha = \frac{\sinh \gamma_\alpha}{e^{-\delta} \cosh \beta_\alpha + \cosh \gamma_\alpha}
	.
	\label{eq:ssbg}
\end{equation}
Minimizing the $\omega$ with respect to $\beta_\alpha$ and $\gamma_\alpha$, one gets the system of MFA equations
\begin{equation}
	\beta_\alpha - \xi = - z\nu S_{\bar{\alpha}}
	,\qquad
	\gamma_\alpha - \eta = - zj \sigma_{\bar{\alpha}}
	,
	\label{eq:mainsys}
\end{equation}
where $\bar{A} = B$, $\bar{B} = A$. 

The Eq. (\ref{eq:mainsys}) should be completed by the charge constraint, $S_A+S_B=2n$. 
To take this condition into account explicitly, we can introduce the charge order parameter $a=(S_A-S_B)/2$,
and write the free energy $f = \omega + \mu n$ as a function of $n$, $a$, and $\sigma_\alpha$ by using the inverse relations for the Eq. (\ref{eq:ssbg}):
\begin{equation}
	e^{2\beta_\alpha}
	= 
	\frac{ 
	\left( S_\alpha e^\delta + G_\alpha \right)^2 - \sigma_\alpha^2 e^{-2\delta} 
	}{
	\left( 1 - S_\alpha \right)^2 - \sigma_\alpha^2
	}
	,\quad
	e^{2\gamma_\alpha}
	= 
	\frac{ 
	\left( \sigma_\alpha e^{-\delta} + G_\alpha \right)^2 - S_\alpha^2 e^{2\delta} 
	}{
	\left( 1 - \sigma_\alpha \right)^2 - S_\alpha^2
	}
	,
	\label{eq:bgss}
\end{equation}
where 
$G_\alpha 
= \left( 1 - S_\alpha^2 - \sigma_\alpha^2 + S_\alpha^2 e^{2\delta} + \sigma_\alpha^2 e^{-2\delta} \right)^{1/2}$.

For the non-ordered (NO) high-temperature solution at $h=0$ we have 
$a = 0$, $\sigma_\alpha = 0$, and the free energy per site has the form
\begin{equation}
	f_{NO}
	= \frac{z}{2} V n^2 
	+ \Delta |n|  
	- \frac{1}{\beta} \ln \left(2 \, \frac{ 1 + g_0 }{ 1 - n^2 }\right)
	+ \frac{|n|}{\beta} \ln \left(\frac{ |n|  + g_0 }{ 1 - |n| }\right)
	,
	\label{eq:fNO}
\end{equation}
where $g_0  =  \left( \left(  1 - n^2 \right) e^{-2\delta}  +  n^2 \right)^{1/2}$.
The expression (\ref{eq:fNO}) for the large positive $\Delta$ and $V=0$ is consistent with the results of~\cite{Capel1966}.
It allows us to find all the thermodynamic properties of the NO phase.   
The entropy, internal energy, and specific heat per site are as follows
\begin{eqnarray}
	s_{NO} 
	&=& 
	\delta \frac{ \left(  1 - |n|  \right)  \left(   g_0 -|n|  \right) }{1 + g_0}
	+ \ln \left(  2 \,\frac{ 1 + g_0 }{ 1 - n^2 }  \right)
	- |n| \ln \left(  \frac{ |n|  + g_0 }{ 1 - |n| }  \right) 
	\label{eq:sNO}
	, \\
	e_{NO} 
	&=& 
	\frac{z}{2} V n^2 +  \Delta \frac{ n^2  + g_0 }{ 1 + g_0 }
	\label{eq:eNO}
	, \\
	c_{NO} 
	&=& 
	\delta^2 \frac{ \left( 1 - n^2 \right)^2  e^{-2\delta} }{ g_0 \left(  1 + g_0  \right)^2 }
	\label{eq:cNO}
	.
\end{eqnarray}
Up to the temperature independent term $\frac{z}{2} V n^2$, the expressions (\ref{eq:fNO}--\ref{eq:cNO}) give the quantities for the ideal system of non-interacting charge (pseudospin) and spin doublets separated in energy by the value $\Delta$. 
At $\Delta=0$, the entropy and internal energy become constant, so the specific heat is zero. 
If $\Delta\neq0$, the specific heat has a maximum at $T \propto |\Delta|$. 
In particular, if $n=0$,
\begin{equation}
	c_{NO}  = \left( \frac{\delta}{2} \right)^2 \cosh^{-2} \frac{\delta}{2}
	,
\end{equation}
and the maximum is at the point $T=|\Delta|/(2x)$, where $x$ is the root of equation $x=\coth x$.

We can write an explicit form of the magnetic susceptibility at $h=0$ in the NO phase. 
Taking that $S_A = S_B = n$ and $\sigma_A = \sigma_B = \sigma$ if $h \neq 0$, 
we eliminate $\xi$ from the system (\ref{eq:mainsys}) and get the equation
\begin{equation}
	\sigma
	=
	\psi ( \eta - zj\sigma  , n)
	,
	\label{eq:sigma}
\end{equation}
where the following notifications have been introduced
\begin{equation}
	\psi(x , n)
	=
	\frac{
	\left( 1-n^2 \right) \sinh x
	}{
	\cosh x + g ( x , n )
	}
	,
\end{equation}
\begin{equation}
	g (x , n) =  \left( \left(  1 - n^2 \right) e^{-2\delta}  +  n^2 \cosh^2 x \right)^{1/2}
	,\quad
	g (0 , n) =  g_0
	. 
	\label{eq:g}
\end{equation}
After the standard calculation, we obtain
\begin{equation}
	\left.  \chi_{NO}  \right|_{h=0}
	= \beta s^2 \left.  \frac{\partial \sigma}{\partial \eta}  \right|_{\eta=0}
	= s^2 \frac{\chi_0(n)}{1 + z \tilde{J} \chi_0(n)}
	,\quad
	\chi_0(n) = \beta \frac{1 - n^2}{1 + g_0}
	,
	\label{eq:chi}
\end{equation}
where $\chi_0(n)$ is the normalized zero-field susceptibility of the ideal system of non-interacting pseudospin and spin doublets. 
Eq.~(\ref{eq:chi}) is consistent with result of~\cite{Capel1967} for the case $n=0$.

The system (\ref{eq:mainsys}) has ferrimagnetic solutions with $\sigma_A + \sigma_B \neq 0$ at $h=0$ \cite{PanovJSNM2016}, 
which are a consequence of the MFA and do not arise in the MC simulations. 
Due to the short-range character of exchange interaction in the model,  
these solutions can manifest itself as a mixture of antiferromagnetic and paramagnetic phases. 
The underestimation of paramagnetic response is a systematic error of MFA in these cases. 
Hereafter, we consider only antiferromagnetic type solutions with $\sigma_A = - \sigma_B = \sigma$ at $h=0$. 
In this case $\gamma_A = -\gamma_B$, as it follows from Eq. (\ref{eq:mainsys}). 
The sublattice magnetizations $\sigma_\alpha$ are monotonic functions of molecular fields $\gamma_\alpha$ in accordance with Eq. (\ref{eq:ssbg}), hence only the case $\beta_A = \pm \beta_B$ is possible for $\sigma \neq 0$. 
It means that if $n \neq 0$ only pure AFM solutions with $a=0$, $\sigma \neq 0$ and pure CO solutions with $a\neq0$, $\sigma = 0$ exist. 
The case $n=0$ should be treated separately, as it provides an opportunity for frustrated states, when the CO and AFM phases  become degenerate.

The thermodynamic properties of the AFM and CO phases assume knowledge of the roots for the Eq. (\ref{eq:mainsys}) and can be calculated numerically.
Besides, we can find analytically the equations for the second-order transition temperatures and for critical points. 

For the AFM phase we employ the condition $\partial^2 f/\partial\sigma^2 = 0$ at $\sigma=0$ that gives
\begin{equation}
	\left. \frac{\partial\gamma_\alpha}{\partial\sigma_\alpha} \right|_{\sigma_\alpha=0} = zj
	.
\end{equation}
With accounting for Eq. (\ref{eq:bgss}) we get the equation for the temperature of the NO-AFM transition 
\begin{equation}
	\left( 1-n^2 \right) zj = 1+g_0
	.
	\label{eq:ct2AFM}
\end{equation}
In particular, for $\Delta \rightarrow +\infty$ we obtain
\begin{equation}
	T_{AFM} = \left( 1-|n| \right) z\tilde{J}
	,
\end{equation}
that coincides with the results of~\cite{Blume1971}.
Substituting Eq. (\ref{eq:ct2AFM}) into Eq. (\ref{eq:chi}) we find the susceptibility in transition point, $\chi_{NO} = s^2/(2z\tilde{J})$.

We use the equation $\partial^4 f/\partial\sigma^4 = 0$ on a coexistence curve to find the critical point that separates the first and second type transitions. 
After some manipulations it reduces to an equation 
\begin{equation}
	g_0^2 - 2 g_0 - 3 n^2 = 0
	.
\end{equation}
With accounting for Eq. (\ref{eq:ct2AFM}) we get the critical point location in the form
\begin{equation}
	T_{c1} 
	= z\tilde{J} \frac{ 1-n^2 }{2 + \sqrt{1+3n^2}}
	,\quad
	\frac{\Delta_{c1}}{T_{c1}} 
	= \frac{1}{2} \ln \frac{ 1-n^2 }{2\left(1+n^2 + \sqrt{1+3n^2}\right)}
	.
\end{equation}
In particular, for  $n=0$, $T_{c1} = z\tilde{J}/3$, $\Delta_{c1} / T_{c1} = - \ln 2$.  
The value of $T_{c1}$ coincides with the results of~\cite{Capel1966,Blume1971},  
but the value of $\Delta_{c1}$ is two times less in our model for this case.

The zero-field susceptibility in the AFM phase has the form
\begin{equation}
	\left.  \chi_{AFM}  \right|_{h=0}
	= s^2 \frac{\beta \psi'(zj\sigma , n)}{1 + zj \psi'(zj\sigma , n)}
	,\quad
\end{equation}
where
\begin{equation}
	\psi'(x , n) = 
	\frac{\left( 1 - n^2 \right) \left( g(x , n) + g_0^2 \cosh x \right)}{g(x , n) \left( \cosh x + g(x , n) \right)^2}
	.
	\label{eq:chiAFM}
\end{equation}
The AFM order parameter $\sigma$ at $h=0$ can be found by solution of equation
\begin{equation}
	\sigma = \psi(zj\sigma , n)
	.
	\label{eq:sigma2}
\end{equation}

Similarly, for the CO phase the condition $\partial^2 f/\partial a^2 = 0$ at $a=0$ gives
\begin{equation}
	\frac{1}{2} 
	\left. 
	\left( 
	\frac{\partial\gamma_A}{\partial a} - \frac{\partial\gamma_B}{\partial a} 
	\right) 
	\right|_{\sigma_\alpha=0} = z\nu
	,
\end{equation}
and for the temperature of the NO-CO transition we get the equation
\begin{equation}
	\left( 1-n^2 \right) z\nu = 1+g_0^{-1}
	.
	\label{eq:ct2CO}
\end{equation}
In particular, for $\Delta \rightarrow -\infty$ we obtain
\begin{equation}
	T_{CO} = \left( 1-n^2 \right) zV
	.
\end{equation}
that coincides with the results of~\cite{Saito1981}.
The equation for the critical point in CO phase is more complicated,
\begin{equation}
	2(1+3n^2) g_0^3 - g_0^2 -6n^2 g_0 + 3 n^2 = 0
	,
\end{equation}
but for $n=0$ it gives $T_{c2} = zV/3$, $\Delta_{c2} / T_{c2} = \ln 2$.

In the CO phase, the zero-field susceptibility has the form
\begin{equation}
	\left.  \chi_{CO}  \right|_{h=0}
	= s^2 
	\frac{ 
	\frac{1}{2} \left( \chi_0(n+a) + \chi_0(n-a) \right)
	}{
	1 + \frac{1}{2} z \tilde{J} \left(\chi_0(n+a) + \chi_0(n-a)\right)
	}
	,
\end{equation}
where the CO parameter satisfies the equation
\begin{equation}
	a 
	= \frac{1}{2z\nu} \ln 
	\left(  
	\frac{\left(n+a+g(0 , n+a)\right) \left(1-n+a\right)}{\left(n-a+g(0 , n-a)\right) \left(1-n-a\right)}  
	\right)
	.
\end{equation}

A general formula for the zero-field susceptibility that combines cases of the NO, AFM and CO phases is given by
\begin{equation}
	\chi
	= s^2 
	\frac{ 
	\frac{1}{2} \beta \left( \psi'(zj\sigma , n+a) + \psi'(zj\sigma , n-a) \right)
	}{
	1 + \frac{1}{2} z j \left(\psi'(zj\sigma , n+a) + \psi'(zj\sigma , n-a)\right)
	}
	,
\end{equation}
where $\sigma$ is the AFM order parameter and $a$ is the CO parameter.

\begin{figure}
\centering
  \includegraphics[width=0.8\textwidth]{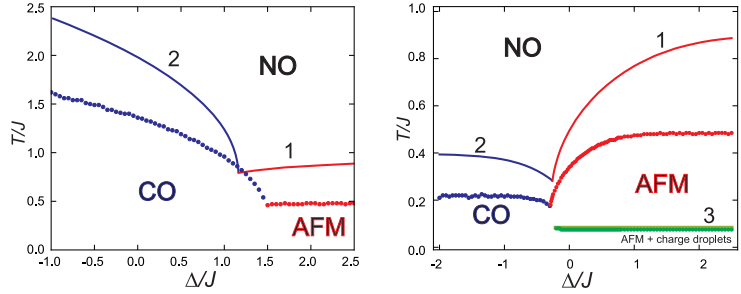}
\caption{
(color online) 
Left panel: the weak exchange case: $n=0.1$, $\tilde{J}=0.25$, $V=1$; 
right panel: the strong exchange case: $n=0.1$, $\tilde{J}=0.25$, $V=0.1$.
Solid circles denote the MC critical temperatures.
Solid lines 1, 2 and 3 show the MFA values of critical temperature given by (\ref{eq:ct2AFM}), (\ref{eq:ct2CO}) and (\ref{eq:Tps}).
}
\label{fig:pd-01-10}      
\end{figure}

For numerical simulation, a high-performance parallel computing program was implemented using the classical MC method. 
The results of the MC calculations are shown in Fig.\ref{fig:pd-01-10}.  
The peak position on the temperature dependence of the specific heat approximately (because of the finite size of the system) corresponds to the transition temperature from non-ordered to ordered state. 
These points are shown by solid circles. 
The MFA transition temperatures (\ref{eq:ct2AFM}) and (\ref{eq:ct2CO}) are shown by solid lines. 
The Fig.\ref{fig:pd-01-10} clearly demonstrates typical, slightly less than twice, overestimation of the critical temperature value by MFA.

\begin{figure}
\centering
  \includegraphics[width=\textwidth]{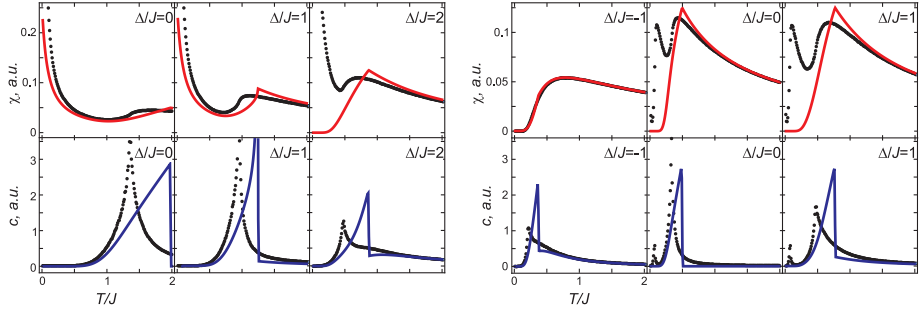}
\caption{
(color online) 
The susceptibility and the specific heat obtained by the MFA (solid lines) and by the MC calculations (solid circles). 
Left panel:  the weak exchange case: $n=0.1$, $\tilde{J}=0.25$, $V=1$; 
right panel: the strong exchange case: $n=0.1$, $\tilde{J}=0.25$, $V=0.1$.
}
\label{fig:susc-spheat}      
\end{figure}

We compare the results for the susceptibility and the specific heat obtained by the MFA and by the MC calculations in Fig.\ref{fig:susc-spheat}.
The analytical MFA dependencies show qualitative match with the numerically calculated ones, and even the quantitative agreement of the results for the high temperature region. 
The main discrepancies are caused by difference in the critical temperature and by systematic inaccuracies of MFA for the description of the critical fluctuations and paramagnetic response at low temperatures.

%%%%%%%%%%%%%%%%%%%%%%%%%%%%%%%%%%%%%%%%%%%%%%%%%%%%%%%%%%%%%%%%%%%%%%%%%%%%%%%

\section{Critical temperature of the spin-charge separation}

The temperature dependencies of the specific heat  
in the strong exchange limit at positive $\Delta$ exhibit two successive phase transitions. 
A direct exploration of the system state shows that first transition is the AFM ordering.
With lowering temperature in the spin subsystem diluted by randomly distributed charged impurities, 
the condensation of impurities in the charge droplets occurs. 
It means that the AFM phase in the strong exchange limit is unstable with respect to macroscopic separation   
of the charge and spin subsystems. 
At this point, the AFM matrix pushes out the charges to minimize the surface energy associated with the impurities.  
Note 
that in the weak exchange limit the charged impurities remain distributed randomly over the AFM matrix up to $T=0$, and also the charged impurities remain distributed randomly in the CO phase, as for the near-neighbor interaction the energies of all possible distributions of extra charges over the CO matrix are equal.

To describe the thermodynamic properties of inhomogeneous state we use the model developed in 
\cite{Kapcia2012,Kapcia2013,Kapcia2015} 
for the macroscopic PS states in electronic systems. 
Assuming the coexistence of two macroscopic homogeneous phases, labeled as 1 and 2, we write the free energy of the PS state per site in the form 
\begin{equation}
	f_{PS} = m \, f_1(n_1)+(1-m) \, f_2(n_2)
	,
\end{equation}
where $m$ is a fraction of the system with a density $n_1$, $1-m$ is a fraction with density $n_2$, 
so that $m \, n_1 + (1-m) \, n_2 = n$. 
In our case, one phase consists of charges (C), and another one is the pure spin AFM phase, 
hence $n_1=\mathop{\mathrm{sgn}} n$, $n_2=0$ and $m=|n|$. 
The transition point is defined by the equation
\begin{equation}
	|n| \, f_C(1) + (1-|n|) \, f_{AFM}(0) = f_{AFM}(n)
	.
	\label{eq:Tps0}
\end{equation}
The free energy of charges is $f_C(1) = 2V + \Delta$. 
The free energy of the AFM phase can be written as
\begin{eqnarray}
	f_{AFM}(n) 
	&=& 
	\frac{z}{2} \left(  V n^2 + \tilde{J} \sigma^2  \right) +  |n| \Delta 
	- \frac{1}{\beta} \ln \left( 2 \, \frac{ \cosh(zj\sigma) + g(zj\sigma , n) }{1 - n^2} \right)
	+{}
	\nonumber
	\\
	&&
	{}+ \frac{|n|}{\beta} \ln \left( \frac{ |n| \cosh(zj\sigma) + g(zj\sigma , n) }{1 - |n|} \right)
	.
\end{eqnarray}
We consider $\Delta>0$ and substantially low temperatures, so that $\delta \gg 1$ and $j \gg 1$. 
In this approximation, with accounting of Eq. (\ref{eq:sigma2}) we obtain $|\sigma| = 1 -|n|$.  
Finally, the equation (\ref{eq:Tps0}) gives the following expression for the PS critical temperature:
\begin{equation}
	T_{PS} = \frac{|n| \big( 1-|n| \big)}{ |n| \ln |n| +(1-|n|) \ln (1-|n|) } \frac{z(V - \tilde{J})}{2}
	.
	\label{eq:Tps}
\end{equation}

The PS in the pseudospin $S=1$ system is a well-known result of the Blume--Emery--Griffiths model~\cite{Blume1971}, 
where the PS curve in the $T{-}n$ plane was obtained by numerically solving the MFA equations. 
The parameter $\Delta$ in the Blume--Emery--Griffiths model consists of the pseudospin interaction parameters and chemical potential, so it is conjugate to the concentration. 
In our model, the on-site correlation parameter $\Delta$ (or zero-field splitting in terms of~\cite{Capel1967}) and the concentration $n$ are independent, so caution should be taken when comparing the results.
Indeed, the PS in our model exists at $|n|\neq0$ in the strong exchange limit for all $\Delta>0$ and 
the expression (\ref{eq:Tps}) does not depend on $\Delta$.
This agrees with the MC results for $T_{PS}$ in Fig.\ref{fig:pd-01-10}.

\begin{figure}
\centering
  \includegraphics[width=0.5\textwidth]{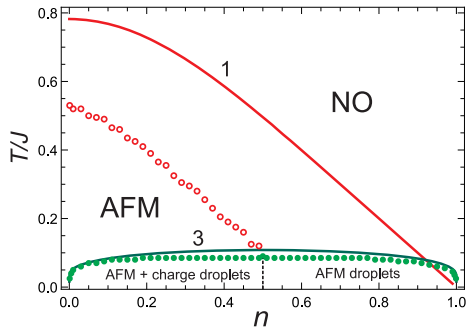}
\caption{
Hollow circles denote the MC results for the maxima of susceptibility due to the AFM ordering, 
and filled circles show the maxima of the specific heat at the PS transition.
Solid lines 1 and 3 show the value of critical temperature given by (\ref{eq:ct2AFM}) and (\ref{eq:Tps}).
}
\label{fig:pd-N}      
\end{figure}

The concentration dependencies of the AFM and PS critical temperatures are shown in Fig.\ref{fig:pd-N}.
Hollow circles denote the MC results for the maxima of susceptibility due to the AFM ordering. 
The AFM transition temperature decreases linearly with the impurity concentration for small $n$, 
which is well known for the site dilute Ising systems~\cite{Ching1976,Landau1977,Lee1979,Velgakis1989,Heuer1992,Souza1992,Sadiq1992,Selke1998,Marques2000}.
For the system with quenched impurities, it continues to decrease to zero at the percolation threshold. 
The PS occurs for annealed and weakly interacting impurities, as in our case. 
Filled circles in Fig.\ref{fig:pd-N} show the maxima of the specific heat at the PS transition. 
Similar phase diagrams was obtained for Blume--Emery--Griffiths model~\cite{Arora1973} and for the site dilute Ising model~\cite{Yaldram1993,Khalil1997}.
Solid line 3 in Fig.\ref{fig:pd-N} denotes the PS critical temperature given by (\ref{eq:Tps}). 
Note that it agrees with the MC results surprisingly well, 
while the MFA dependence 2 for critical temperature of the AFM ordering given by (\ref{eq:ct2AFM}) becomes improper at $|n|>0.5$. 
The concentration dependence similar to our $T_{PS}(n)$ was found numerically for Blume--Capel model~\cite{Butera2018} by intersecting the low-temperature and high-temperature expansions of the free-energy.

\section{Conclusion}

We have addressed a static 2D spin-pseudospin model on a square lattice, 
that generalizes the 2D dilute antiferromagnetic Ising model. 
We compared the analytical MFA results with numerical results of classical MC calculations. 
An analysis of the specific heat and susceptibility obtained by MC method showed that the MFA critical temperatures both for the CO and AFM ordering qualitatively reproduce the numerical results, but systematically give higher values. 
The MC calculations show that the cases of strong exchange and strong charge correlation differ qualitatively. 
In the case of strong charge correlation, one has a frustration in the charge-ordered ground state of the system. 
The homogeneous AFM phase in the strong exchange limit is unstable with respect to PS of the charge and spin subsystems. 
An analytical expression was obtained for PS temperature, and we found that it agrees well with the numerical results.

\bigskip
The work was supported  by the Government of the Russian Federation, Program 02.A03.21.0006 and by the Ministry of Education and Science of the Russian Federation, projects Nos. 2277 and 5719.

\end{document}